\documentclass{article}

\usepackage{PRIMEarxiv}

\usepackage[utf8]{inputenc} 
\usepackage[T1]{fontenc}    
\usepackage{hyperref}       
\usepackage{url}            
\usepackage{booktabs}       
\usepackage{amsfonts}       
\usepackage{nicefrac}       
\usepackage{microtype}      
\usepackage{lipsum}
\usepackage{fancyhdr}       
\usepackage{graphicx}       
\graphicspath{{media/}}     
\usepackage{enumerate}
\usepackage{amsmath}
\usepackage{float}
\usepackage{subcaption}
\newcommand*\samethanks[1][\value{footnote}]{\footnotemark[#1]}

\title{Making Thermodynamic Models of Mixtures Predictive by Machine Learning: Matrix Completion of Pair Interactions
\thanks{\textit{\underline{Published version, cite as follows}}: 
\textbf{F. Jirasek, R. Bamler, S. Fellenz, M. Bortz, M. Kloft, S. Mandt, H. Hasse: Making Thermodynamic Models of Mixtures Predictive by Machine Learning, Chemical Science 13 (2022) 4854-4862. DOI:10.1039/D1SC07210B.}} 
}

\author{
  Fabian Jirasek \\
  Laboratory of Engineering Thermodynamics (LTD) \\
  TU Kaiserslautern \\
  Kaiserslautern, Germany \\
  \texttt{fabian.jirasek@mv.uni-kl.de} \\ 
  \And
  Robert Bamler \\
  Data Science and Machine Learning \\
  University of T\"ubingen \\
  T\"ubingen, Germany \\
  \And
  Sophie Fellenz \\
  Machine Learning Group\\
  TU Kaiserslautern\\
  Kaiserslautern, Germany\\
  \And
  Michael Bortz \\
  Fraunhofer Institute for\\ 
	Industrial Mathematics (ITWM)\\
  Kaiserslautern, Germany\\
  \And
  Marius Kloft \\
  Machine Learning Group\\
  TU Kaiserslautern\\
  Kaiserslautern, Germany\
  \And
  Stephan Mandt\thanks{These authors jointly supervised this work.} \\
  Department of Computer Science\\
  University of California, Irvine\\
  Irvine, USA\\
	\And
	Hans Hasse\samethanks \\
  Laboratory of Engineering Thermodynamics (LTD) \\
  TU Kaiserslautern \\
  Kaiserslautern, Germany \\
}

\begin{document}
\maketitle

\begin{abstract}
Predictive models of thermodynamic properties of mixtures are paramount in chemical engineering and chemistry. Classical thermodynamic models are successful in generalizing over (continuous) conditions like temperature and concentration. On the other hand, matrix completion methods (MCMs) from machine learning successfully generalize over (discrete) binary systems; these MCMs can make predictions without any data for a given binary system by implicitly learning commonalities across systems. In the present work, we combine the strengths from both worlds in a hybrid approach. The underlying idea is to predict the \textit{pair-interaction energies}, as they are used in basically all physical models of liquid mixtures, by an MCM. As an example, we embed an MCM into UNIQUAC, a widely-used physical model for the Gibbs excess energy. We train the resulting hybrid model in a Bayesian machine-learning framework on experimental data for activity coefficients in binary systems of 1,146 components from the Dortmund Data Bank. We thereby obtain, for the first time, a \textit{complete} set of UNIQUAC parameters for all binary systems of these components, which allows us to predict, in principle, activity coefficients at arbitrary temperature and composition for any combination of these components, not only for binary but also for multi-component systems. The hybrid model even outperforms the best available physical model for predicting activity coefficients, the modified UNIFAC (Dortmund) model.
\end{abstract}


\section{Introduction}
Information on thermodynamic properties of mixtures is of crucial importance in chemical engineering and chemistry. However, providing this information is hampered by a combinatorial problem: the number $N$ of known components is increasing rapidly (it is presently in the order of $N = 10^8$, counted by entries in the CAS Registry~\cite{CAS_Registry}); if only binary mixtures are considered, the number of mixtures that can be formed from these components already goes with $N^2$. Even if only technically relevant components and their mixtures are considered, the numbers are still extremely high. Consequently, experimental data on thermodynamic properties are available only for a small fraction of the relevant mixtures, especially as the corresponding experimental investigations are tedious. Therefore, methods for the prediction of thermodynamic properties of mixtures are essential in practice.

The present work is focused on thermodynamic properties of fluid mixtures. Physical models for describing these properties are generally based on the concept of \textit{pair interactions}, which are commonly described by \textit{pair-interaction energies}. All types of pair interactions in a multicomponent mixture can be investigated by studying the pure components (for the like interactions) and the binary subsystems (for the unlike interactions). Therefore, physical models for thermodynamic properties of fluid mixtures are generally developed using data on pure components and binary mixtures; based on this, they can also be used to predict properties of multi-component mixtures.  
In the application of models of thermodynamic properties, their predictive capabilities are of prime importance. To assess them, the relation of the data used in the model development (the \textit{training data}) to the data to be predicted is important. Regarding thermodynamic properties of mixtures, we distinguish two categories of predictions here: 
\begin{enumerate}[(i)]
\item those, in which for a given system (a fixed set of components) only the conditions are changed compared to the training data (e.g., temperature, pressure, or concentration of the components); we refer to this as \textit{generalization over conditions}. 
\item those, in which the system itself is changed, i.e., was not included in the training data; we refer to this as \textit{generalization over systems}. 
\end{enumerate}

While (i) involves \textit{continuous} variables, (ii) is \textit{discrete}. 
In the nomenclature of the present paper, we follow the common usage, which says that \textit{'mixture'} can have two meanings: in the first, it simply designates the combination of specific components (regardless of their concentrations, e.g., 'water + ethanol'), in the second, also the concentrations are included in the term (e.g., 'water + ethanol with $x_\mathrm{water}=0.1$~mol/mol'). In cases where this ambiguity can lead to misunderstandings, we use the term \textit{'system'} instead of 'mixture' when we refer to the first case.

The most common types of thermodynamic models of mixtures are models of the Gibbs excess energy $G^\mathrm{E}$, such as UNIQUAC~\cite{abrams1975,maurer1987} or NRTL~\cite{renon1986}, and equations of state (EoS)~\cite{poling_2001}. These models excel in generalizing over conditions. By virtue of the underlying thermodynamic theory, they can also be used for \textit{generalizing over properties}, i.e., if trained on data for a specific property, they can be used to predict a different but related property; this is an important issue, which, however, is not in the focus of the present work. 

In their basic form, $G^\mathrm{E}$ models and EoS are only partially useful for generalizing over systems: for the reasons given above, they are usually trained on data for binary systems, but can then be used for modeling those binary systems only for which data were available for training. In contrast, predictions of properties of unstudied multicomponent systems with $G^\mathrm{E}$ models and EoS often turn out to be sufficiently accurate, under the condition that the model was trained on data for \textit{all constituent binary subsystems}~\cite{carlson1942}.  

To overcome the lack of generalization over (binary) systems, group-contribution (GC) approaches like UNIFAC~\cite{fredenslund_1975,fredenslund_1977,fredenslund1989}, modified UNIFAC (Dortmund)~\cite{weidlich_1987,constantinescu_2016}, and the Predictive Soave-Redlich-Kwong (PSRK) EoS~\cite{holderbaum1991,horstmann2005} have been 
developed. They are based on the idea that components can be split into chemical groups and, instead of considering pair interactions between components, pair interactions between these groups are modeled. As the number of relevant chemical groups is comparatively small (in the order of 100), the combinatorial problem described above thereby becomes tractable. GC methods contain group-interaction parameters that are usually trained on data for binary systems containing the pertinent groups. They can then be used for predicting the properties of systems for which no data are available. 

Unfortunately, the applicability of these GC methods is still hampered by incomplete group-interaction parameter tables. This is due to both the elaborate procedure of fitting new group-interaction parameters, and the fact that often not enough pertinent experimental data are available for a meaningful fit. The problems resulting from an inadequate database also lead to poor performances of GC methods when applied to systems with components that contain groups for which only a few training data points are available. 
The most prominent alternatives to GC methods are quantum-chemical approaches, namely COSMO-RS~\cite{klamt_1995,klamt1998refinement}. In principle, COSMO-RS enables predictions for any system based on quantum-chemical calculations, which are, however, computationally costly and not trivial for complex components. Furthermore, also these methods are tuned on experimental data, but the number of parameters is typically low and common users do not change them~\cite{klamt2010}. Compared to GC methods, quantum-chemical methods are often found to be somewhat less accurate in the prediction of thermodynamic properties~\cite{grensemann2005}.  

We have recently introduced a completely new approach for predicting thermodynamic properties of unstudied binary systems~\cite{jirasek_2020, jirasek2020c, Damay2021, Jirasek2021Perspective}. This approach is based on employing matrix completion methods (MCMs) from machine learning (ML), where the MCMs are prominently associated with recommender systems~\cite{koren2009matrix, takacs2008investigation}. E.g., for movie recommendations, these methods can implicitly learn and quantify similarities among users and similarities among movies by observing interaction patterns (ratings or clicks) between them, allowing to predict preference scores for unseen pairs of users and movies. In a similar spirit, our previous work~\cite{jirasek_2020,jirasek2020c} employed MCMs for predicting activity coefficients $\gamma_{ij}^\infty$ of solutes $i$ in pure solvents $j$ at infinite dilution and constant temperature, where the MCM learns similarities among the solutes and among the solvents. Applying the MCM approach for predicting thermodynamic properties is based on the fact that properties of binary mixtures at constant conditions, such as isothermal $\gamma_{ij}^\infty$, can be stored in matrices, where the rows and columns represent the components $i$ and $j$ that make up the mixtures. As these matrices are only sparsely occupied by experimental data, the prediction of the unobserved entries constitutes a matrix completion problem. In our previous work~\cite{jirasek_2020,jirasek2020c}, we employed a matrix factorization following Eq.~\ref{Eq_MCM}:
\begin{equation}
    \gamma_{ij}^\infty = f(\theta_i \cdot \beta_j)
    \label{Eq_MCM}
\end{equation}
The value of $\gamma_{ij}^\infty$ is thereby modeled by the dot product of two vectors $\theta_i$ and~$\beta_j$, which contain the so-called latent features and capture properties of the solute $i$ and the solvent~$j$, respectively and which are inferred from the sparse available data on the mixture property $\gamma_{ij}^\infty$. The function~$f$ was chosen to be the exponential function, taking into account that the $\gamma_{ij}^\infty$ are by definition non-negative and span a wide range of values. Moreover, this choice is also a physical one since the excess chemical potential $\mu_{ij}^{E,\infty}$ of the solute $i$ at infinite dilution in the solvent $j$ depends linearly on $\ln\gamma_{ij}^\infty$.

The basic idea behind this MCM approach is that \emph{information on the pure components} (as quantified by $\theta_i$ and $\beta_j$) is implicitly contained in the \emph{mixture data}, i.e., in the way the components interact with each other, which manifests itself in the observable mixture property $\gamma_{ij}^\infty$. This information is captured by the MCM and stored in the latent features $\theta_i$ and $\beta_j$ of the components $i$ and $j$. 
Using Eq.~\ref{Eq_MCM} in the trained model, the inferred latent features allow predicting $\gamma_{ij}^\infty$ also for previously unstudied combinations $i$--$j$. As demonstrated in our previous work~\cite{jirasek_2020,jirasek2020c}, this approach even outperforms the present state-of-the-art method for predicting activity coefficients, namely the modified UNIFAC (Dortmund) model~\cite{weidlich_1987,constantinescu_2016}, if $\gamma_{ij}^\infty$ at 298~K are considered. We have also extended the approach to modeling the dependence of $\gamma^\infty_{ij}$ on the temperature $T$, by simply exploiting the fact that this dependence can be well described by $\gamma^\infty_{ij}(T) = A_{ij} + B_{ij}/T$ with system-specific but temperature-independent parameters $A_{ij}$ and $B_{ij}$ in many cases~\cite{Damay2021}.

In the present work, we further expand the MCM approach by combining it with the physical modeling of thermodynamic properties of mixtures. We thereby exploit the fact that all physical models of mixtures are based on the idea of pair interactions, described by pair-interaction energies as critical parameters. We propose to predict these pair-interaction energies by an MCM to obtain a new \textit{hybrid} concept for the prediction of properties of mixtures. The feasibility and the merits of this widely applicable approach are demonstrated by using the well-known $G^\mathrm{E}$ lattice model UNIQUAC~\cite{abrams1975,maurer1987} as an example. 

En passant, we come back to the derivation of the UNIQUAC equation, in which two asymmetric pair-interaction parameters $\Delta U_{ij} \neq \Delta U_{ji}$ were introduced instead of using one symmetric pair-interaction energy $U_{ij} = U_{ji}$ as it follows from the lattice theory. The two adjustable parameters $\Delta U_{ij}$ and $\Delta U_{ji}$ were introduced as a workaround to get more flexibility for fitting binary phase equilibrium data. We show that this workaround is not necessary and base our considerations on the physical symmetric interaction energies $U_{ij} = U_{ji}$, explicitly including those for the like interactions~$U_{ii}$. 

The resulting new model, which we call 'MCM-UNIQUAC' in the following, combines the capabilities of the MCM regarding the generalization over binary systems with those of the physical model UNIQUAC regarding the generalization over conditions. It is a hybrid method that takes advantage of the strengths of both worlds; in particular, it also enables predictions of properties of multicomponent systems by virtue of the physics behind UNIQUAC, a feature a data-driven MCM cannot provide.

\section{Development of MCM-UNIQUAC}
\begin{figure*}
    \centering
    \includegraphics[width=0.85\textwidth]{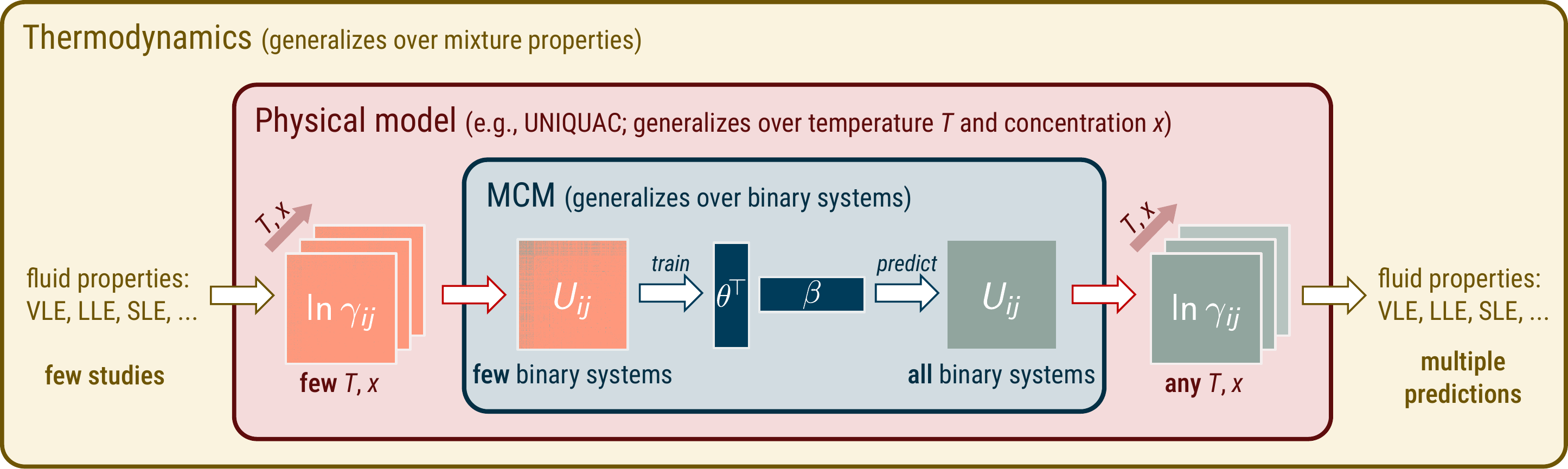}
    \caption{Illustration of embedding an MCM into a physical model of mixtures (here: the lattice model UNIQUAC). Blue part: application of an MCM to pair-interaction energies $U_{ij}$. Red part: the physical model relates $U_{ij}$ to temperature- and concentration-dependent activity coefficients $\ln\gamma_{ij}$. Yellow part: $\ln\gamma_{ij}$ are directly related to observable mixture properties (e.g., vapor-liquid equilibria (VLE)~\cite{fredenslund2012vapor}, liquid-liquid equilibria (LLE)~\cite{anderson1978application}, solid-liquid equilibria (SLE)~\cite{gmehling1978solid}) by thermodynamic laws.}
    \label{fig:overview}
\end{figure*}

The idea behind our approach is shown in Figure~\ref{fig:overview}. At its heart, an MCM is used to predict pair-interaction energies $U_{ij}$ of a physical model of mixtures, which is UNIQUAC in the present work. The physical model relates the pair-interaction energies to temperature- and concentration-dependent properties of the binary mixtures of the components $i$ and $j$. This could, in principle, be any property of interest. In the present work, we have chosen a fundamental thermodynamic property, the activity coefficient $\gamma_{ij}$ of component $i$ in a binary mixture with component $j$ normalized according to Raoult's law. The activity coefficients, in turn, are directly related to observable mixture properties (e.g., vapor-liquid equilibria (VLE)~\cite{fredenslund2012vapor}, liquid-liquid equilibria (LLE)~\cite{anderson1978application}, and solid-liquid equilibria (SLE)~\cite{gmehling1978solid}) by thermodynamic laws. This enables training the approach on the corresponding thermodynamic data of different types so that many data sources are accessible. The model can be written as: 
\begin{equation}
    \ln\gamma_{ij}(T,x_i) = f_\mathrm{UNIQUAC}(T,x_i,P_i,P_j,U_{ii},U_{jj},U_{ij}) 
    \label{Eq_MCM_UNIQUAC}
\end{equation}
where the function $f_\mathrm{UNIQUAC}$ contains the UNIQUAC equation, which is defined in Eqs.~S.1 - S.7 of the Supporting Information. Here, $T$ is the temperature and $x_i$ is the mole fraction of component $i$ in the binary mixture. 
Eq.~\ref{Eq_MCM_UNIQUAC} contains two types of parameters: first, the geometric pure-component parameters $P_i$ and $P_j$, which are reported for many components (e.g., in the Dortmund Data Bank (DDB)~\cite{DDB_2020}) or can be readily estimated (e.g., with the approach described in connection with the development of the UNIFAC method~\cite{fredenslund_1975}); and second, the pair-interaction energies, where we distinguish the like interactions $U_{ii}$ and $U_{jj}$, which are also pure-component parameters, and the unlike interactions $U_{ij}$ ($i\neq j$), which are binary parameters. It follows from the physical interpretation of $U_{ij}$ as pair-interaction energies that $U_{ij} = U_{ji}$.

Furthermore, it follows from the derivation of UNIQUAC that all model parameters are independent of temperature and concentration. We use this assumption throughout the present work but note that some authors work with temperature-dependent parameters. 
Further, as already mentioned in the introduction, it is common in the literature not to use the symmetric interaction energies $U_{ij} = U_{ji}$ as parameters, but rather the two parameters $\Delta U_{ij}\neq \Delta U_{ji}$, which are calculated by:
\begin{align}
    \Delta U_{ij} = U_{ij} - U_{jj}; \qquad \Delta U_{ji} = U_{ij} - U_{ii}
    \label{eq:delta-u-u}
\end{align}
The correlations between $\Delta U_{ij}$ and $\Delta U_{ji}$ defined by Eq.~\ref{eq:delta-u-u} are usually ignored and the two parameters are 
fitted \textit{independently} for each binary system $i$--$j$ under consideration, simply to increase the flexibility of the model.\footnote[3]{
For a set of $N$ components, there are $N(N-1)/2$ binary systems and, following the common usage, $N(N-1)$ binary UNIQUAC parameters $\Delta U_{ij}$. Using the pair-interaction energies $U_{ij}$ directly as parameters, there are $N(N+1)/2$ parameters, of which $N$ are pure-component parameters. Hereby $N(N-1)>N(N+1)/2$ for $N\geq3$.}
Fitting $\Delta U_{ij}$ and $\Delta U_{ji}$ independently to data for binary systems of $N \geq 3$ components will, in general, lead to results that cannot be reconciled with Eq.~\ref{eq:delta-u-u}, as demonstrated in detail in the Supporting Information. For MCM-UNIQUAC, we only consider the truly physical pair-interaction energies $U_{ij}$.

UNIQUAC allows generalizing (i.e., interpolating and extrapolating) over temperatures and concentrations, but not over mixture components. Applying UNIQUAC to a mixture for which no parameters are available requires at least some mixture data points for determining the pair-interaction parameters. This is a severe restriction, as experimental data are often unavailable, especially for mixtures. To overcome this limitation, we introduce MCM-UNIQUAC to extend UNIQUAC so that it can generalize over temperatures and concentrations \textit{and} over binary systems.

The proposed generalization over binary systems is achieved by an MCM and illustrated in the central (blue) panel in Figure~\ref{fig:overview}.
We thereby model the unlike pair-interaction energies $U_{ij}$ as a sum of dot products:
\begin{align}
    U_{ij} = \theta_i \cdot \beta_j + \theta_j \cdot \beta_i, \qquad i \neq j
    \label{eq:mcm-energies}
\end{align}
where $\theta_i$ and $\beta_i$ as well as $\theta_j$ and $\beta_j$ are the feature vectors for the components $i$ and $j$, respectively, and the right-hand side of Eq.~\ref{eq:mcm-energies} is constructed in such a way that the physical constraint $U_{ij} = U_{ji}$ $\forall$ $i,j$ is enforced, resulting in a \textit{symmetric} matrix $\mathbf{U}$ of the pair-interaction energies. Besides the feature vectors, the like pair-interaction energies $U_{ii}$ and $U_{jj}$ are considered as parameters of MCM-UNIQUAC. Note that all parameters of MCM-UNIQUAC are \textit{component-specific} ($P_i,P_j,\theta_i,\beta_i,U_{ii},U_{jj}$), but they are inferred from \textit{mixture data}. After fitting these parameters, a \textit{complete set} of pair-interaction energies for all conceivable binary combinations $i$--$j$ of all considered components is obtained from Eq.~\ref{eq:mcm-energies}.

MCM-UNIQUAC was trained end-to-end on a set of measured logarithmic temperature- and con\-cen\-tra\-tion-dependent activity coefficients in binary mixtures $\ln\gamma_{ij}$ (red panel of Figure~\ref{fig:overview}). We used data from the DDB~\cite{DDB_2020} here; in specific, we used $\ln\gamma_{ij}$ derived from binary vapor-liquid equilibrium (VLE) data using the extended Raoult's law, cf. Eq.~S.14 in the Supporting Information, which we augmented with temperature-dependent data on binary activity coefficients at infinite dilution $\ln\gamma_{ij}^\infty$.
In total, we obtained a set of 363,181 experimental data points for $\ln\gamma_{ij}$ for 12,199 different binary systems $i$--$j$ involving 1,146 distinct components $i,j$ at varying concentrations and temperatures ranging from 183~K to 638~K. The considered $N=1{,}146$ components result in $N(N-1)/2 = 656{,}085$ possible different binary systems. Experimental data are only available for $12,199$ of these systems, i.e., data are available for less than 2\% of all systems and, consequently, only less than 2\% of these systems can be modeled with UNIQUAC in the conventional way. More details on the data set are given in the Supporting Information.

The systems for which data are available were divided into three sets: 80\% were used for training the model (training set), 10\% were used for setting the model's hyperparameters (validation set), and 10\% were used for testing the predictions (test set). We trained our model using the probabilistic programming language Stan\cite{carpenter2017} and resorted to Variational Inference \cite{blei2017variational,zhang2018advances,kucukelbir2017} for performing approximate Bayesian inference. Details on the random data split, the model training (including the source code to run the model in Stan), and the hyperparameter selection are given in the Supporting Information.

After the training, MCM-UNIQUAC can predict temperature- and concentration-dependent activity coefficients in any binary and multicomponent mixture of the considered 1,146 components. The activity coefficients can, in turn, be used for predicting observable mixture properties, such as VLE or other phase equlibria (yellow panel in Figure~\ref{fig:overview}). We demonstrate the predictive capacity of MCM-UNIQUAC by considering the data from the test set in the following section. 

\section{Results and discussion}
\begin{figure*}[h]
    \centering
    \includegraphics[width=0.70\textwidth]{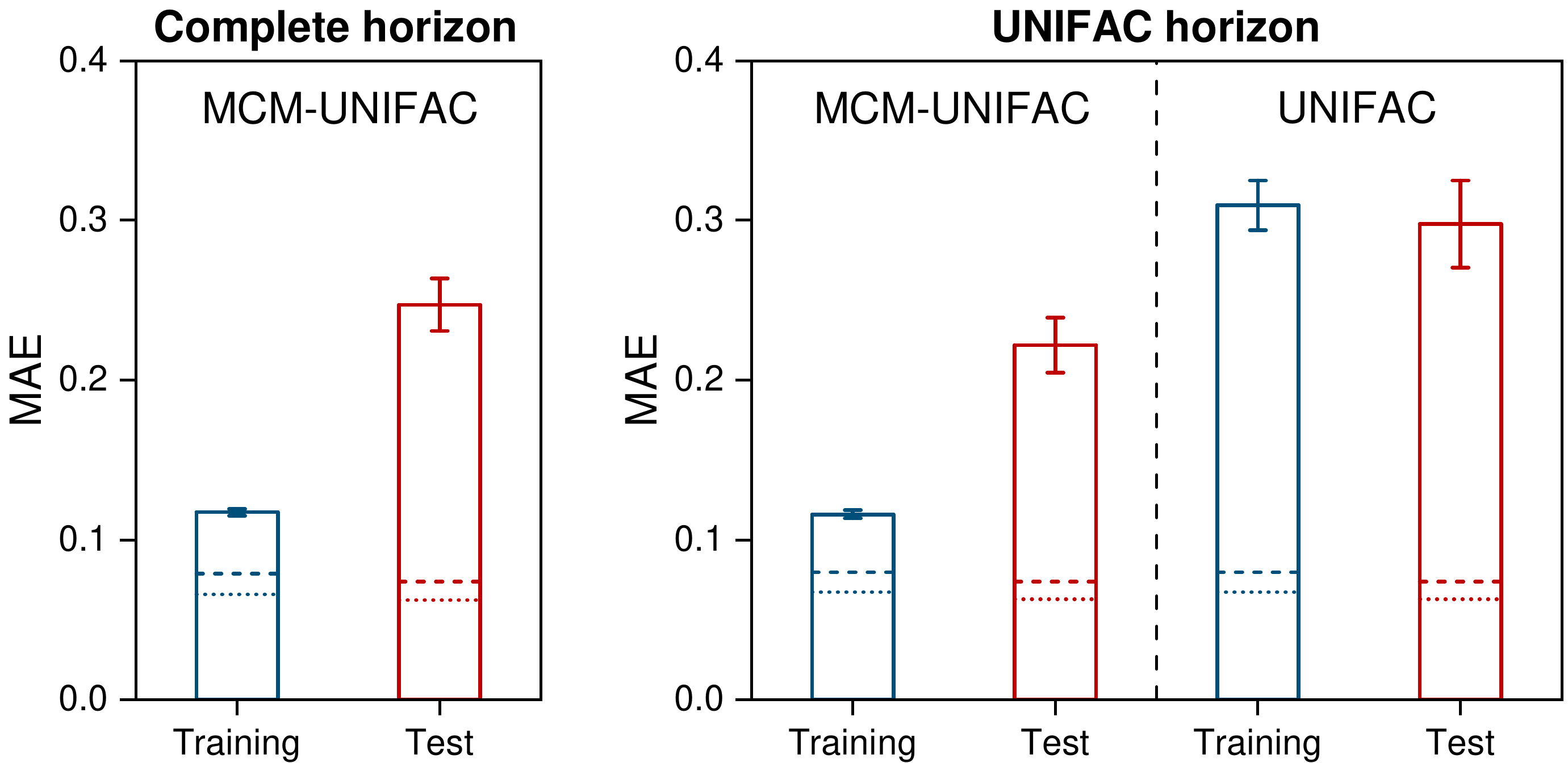}
    \caption{Mean absolute error (MAE) of MCM-UNIQUAC on the training and test set (left) and comparison to UNIFAC based only on those systems that can also be modeled with UNIFAC (right). Bars indicate the results of MCM-UNIQUAC and UNIFAC, and lines denote the baselines obtained by directly fitting UNIQUAC pair-interaction parameters ($\Delta U_{ij}$, dotted) or pair-interaction energies ($U_{ij}$, dashed) to all available data points. Error bars denote standard errors of the means.}
    \label{fig:scores_systems}
\end{figure*}

The results that were obtained with MCM-UNIQUAC are shown as bars in Figure~\ref{fig:scores_systems} (left), where the mean absolute error (MAE) is reported both for the systems from the training and those from the test set; in Figure~S.2 in the Supporting Information, we show the respective results for the mean squared error (MSE). As expected, the error metrics for the test set are larger than those for the training set, but in both cases the overall agreement between the results of MCM-UNIQUAC and the experimental data is remarkable. We demonstrate this in Figure~\ref{fig:scores_systems} (right) by comparison with the best available physical method for the prediction of activity coefficients, the group-contribution model modified UNIFAC (Dortmund)~\cite{weidlich_1987,constantinescu_2016}, which is called 'UNIFAC' in the following for brevity. Unfortunately, in contrast to MCM-UNIQUAC, UNIFAC cannot be applied to all systems for which data are available (denoted as 'complete horizon' in Figure Figure~\ref{fig:scores_systems} (left)) because multiple group-interaction parameters are missing. Hence, for a fair comparison of both methods, only those subsets of the training set and of the test set for which UNIFAC could be applied were used; this 'UNIFAC horizon' covers 7,578 of 9,759 systems from the training set and 961 of 1,220 systems from the test set.

Furthermore, as baselines, the scores obtained for the different sets by directly fitting UNIQUAC parameters to all available data points are marked as lines in Figure~\ref{fig:scores_systems}. We thereby considered two variants: first, UNIQUAC was used in the usual manner by fitting the two binary parameters $\Delta U_{ij}$ and $\Delta U_{ji}$ \textit{individually} to the data for each system $i-j$; this procedure is denoted as 'UNIQUAC~($\Delta U$)' in the following and the respective results are shown as dotted lines in Figure~~\ref{fig:scores_systems}. Second, we fitted the symmetric pair-interaction energies $U_{ij} = U_{ji}$ to all data points; this procedure is denoted as 'UNIQUAC~($U$)' in the following and the respective results are shown as dashed lines in Figure~~\ref{fig:scores_systems}. Note that considering the full set of 12,199 binary systems for which experimental data are available, there are 24,398 parameters in UNIQUAC~($\Delta U$), whereas there are only 13,345 parameters in UNIQUAC~($U$), of which 1,146 are pure-component parameters $U_{ii}$. 

Let us consider the scores of the baselines UNIQUAC~($\Delta U$) and UNIQUAC~($U$) first. They simply indicate how well the data can, in principle, be described with the physical model, i.e., UNIQUAC here. The first astonishing message is that UNIQUAC~($U$) works almost as well as UNIQUAC~($\Delta U$), even though the latter has almost twice as many parameters, cf. above. We have therefore based our hybrid model MCM-UNIQUAC on the more physical UNIQUAC~($U$). 

Second, we observe in Figure~\ref{fig:scores_systems} that MCM-UNIQUAC is in general flexible enough for describing the data well, cf. the relatively small differences between the scores of MCM-UNIQUAC on the training set (blue bars) and the baseline scores (lines). 

Third, by comparing the scores for the test set (red bars), we find that MCM-UNIQUAC clearly outperforms UNIFAC in accurately predicting $\ln\gamma_{ij}$ in both MAE and MSE (cf. Figure S.2 in the Supporting Information). This is particularly remarkable since UNIFAC has been fitted to most of these data points, whereas no data point from the test set was used for training or validation of MCM-UNIQUAC. An alternative representation of the results for the test set is shown in histogram plots in Figures S.3 and S.4 in the Supporting Information, which underpin the superior performance of MCM-UNIQUAC compared to UNIFAC. Similar results are found by considering COSMO-SAC-dsp~\cite{Hsieh2014cosmo,Bell2020cosmo} as baseline, as shown in Figure S.5 in the Supporting Information. Additionally, we discuss the influence of the number of training data points on the performance of MCM-UNIQUAC in Figure S.6 and Table S.1 in the Supporting Information. 

We note that also a version of MCM-UNIQUAC based on UNIQUAC ($\Delta U$), i.e., based on the asymmetric pair-interaction parameters, can be trained if \textit{and only if} an end-to-end training on $\ln\gamma_{ij}$ is performed; the resulting performance is, however, slightly worse than that of the model discussed above as we demonstrate in Figure S.7 in the Supporting Information.

\begin{figure*}
    \centering
    \includegraphics[width=0.35\textwidth]{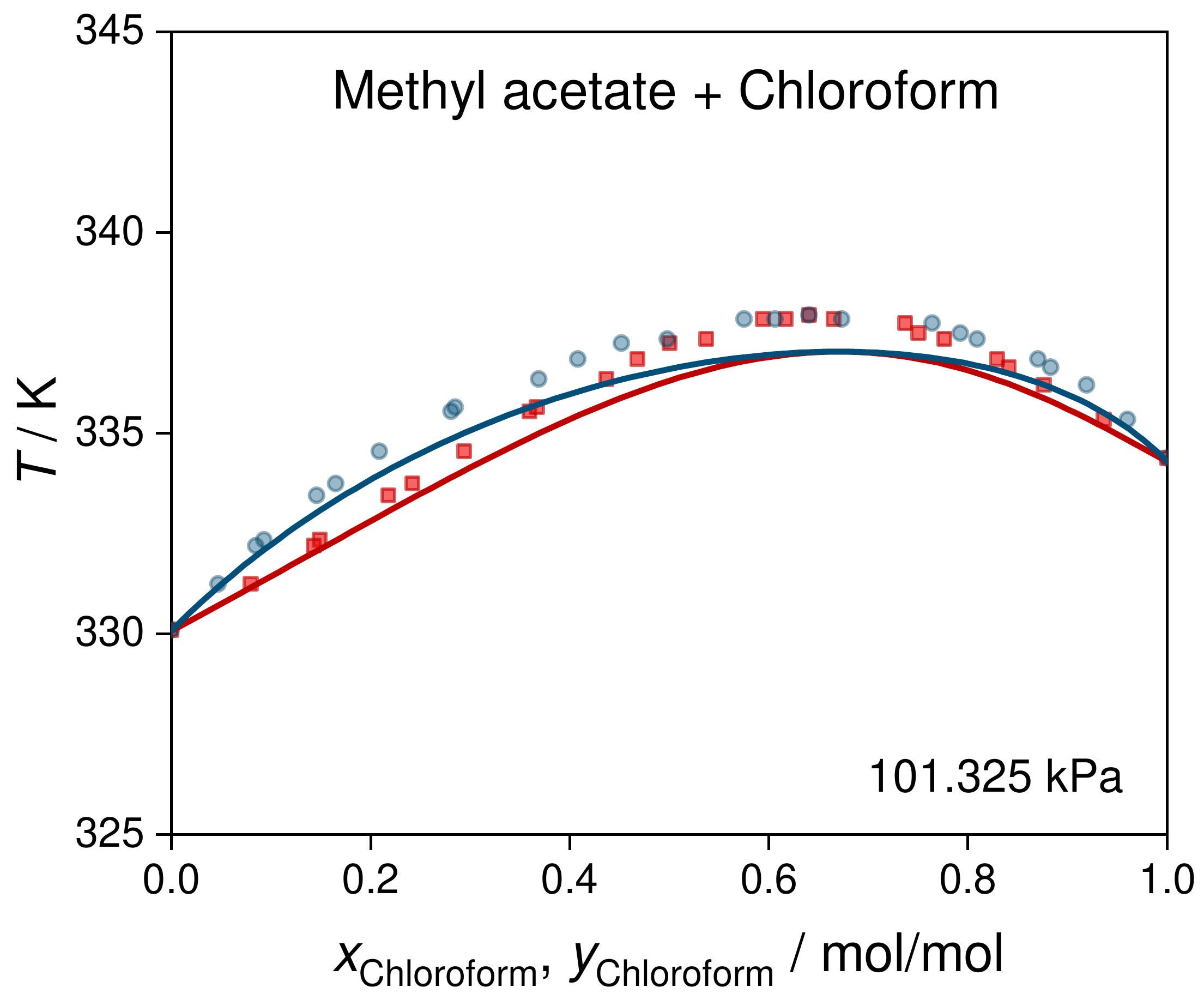}
    \includegraphics[width=0.35\textwidth]{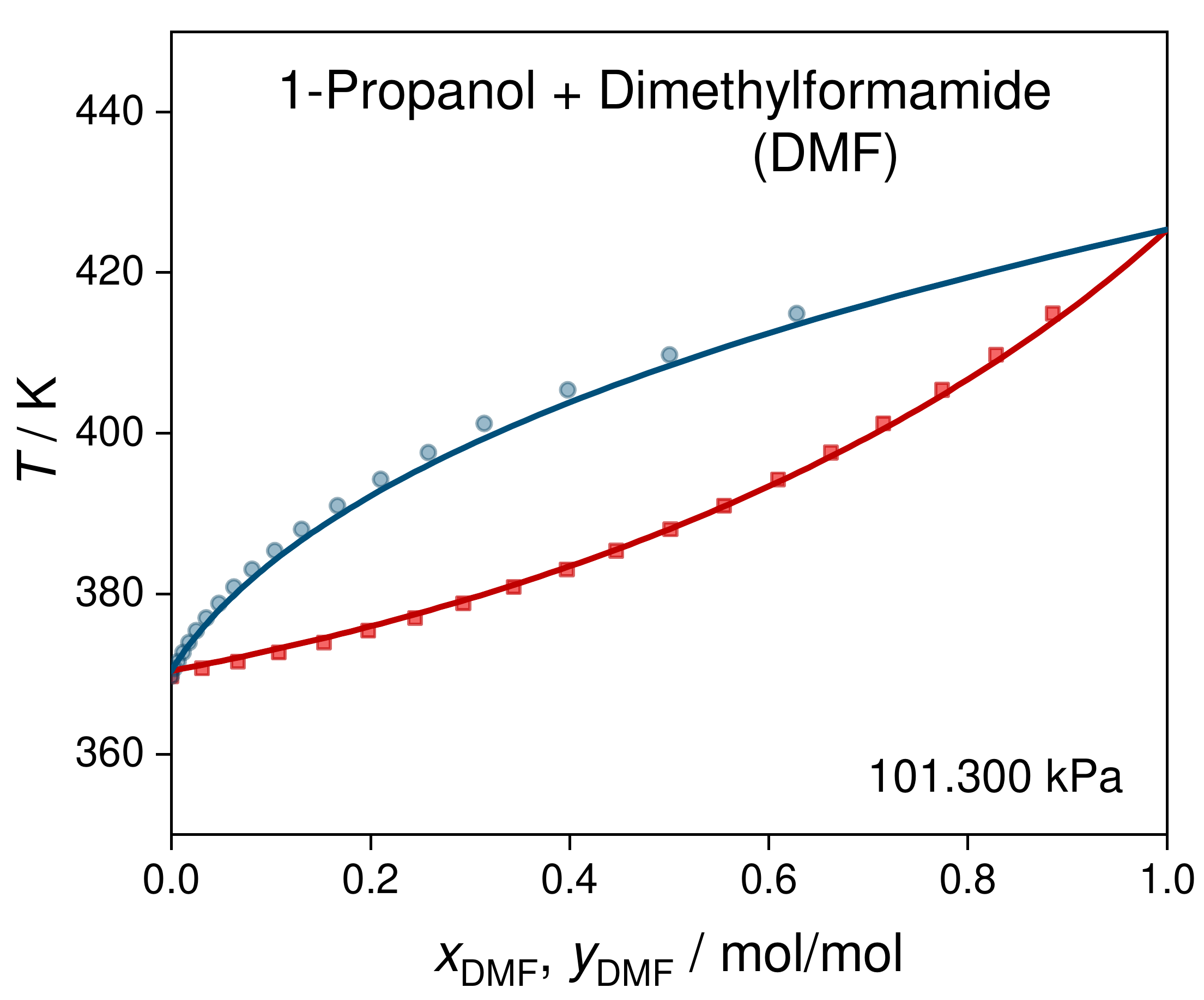}
    \medskip
    \includegraphics[width=0.35\textwidth]{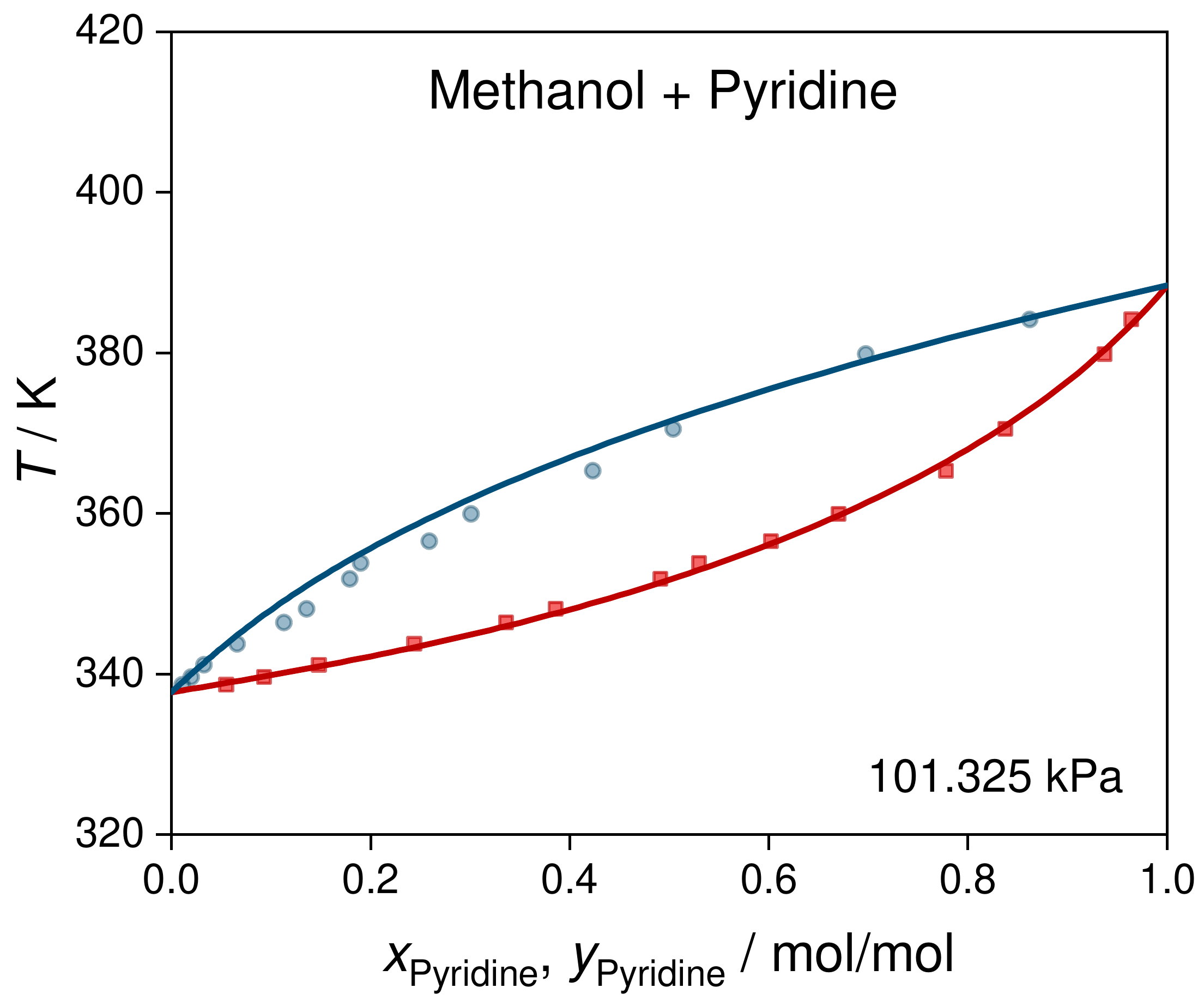}
    \includegraphics[width=0.35\textwidth]{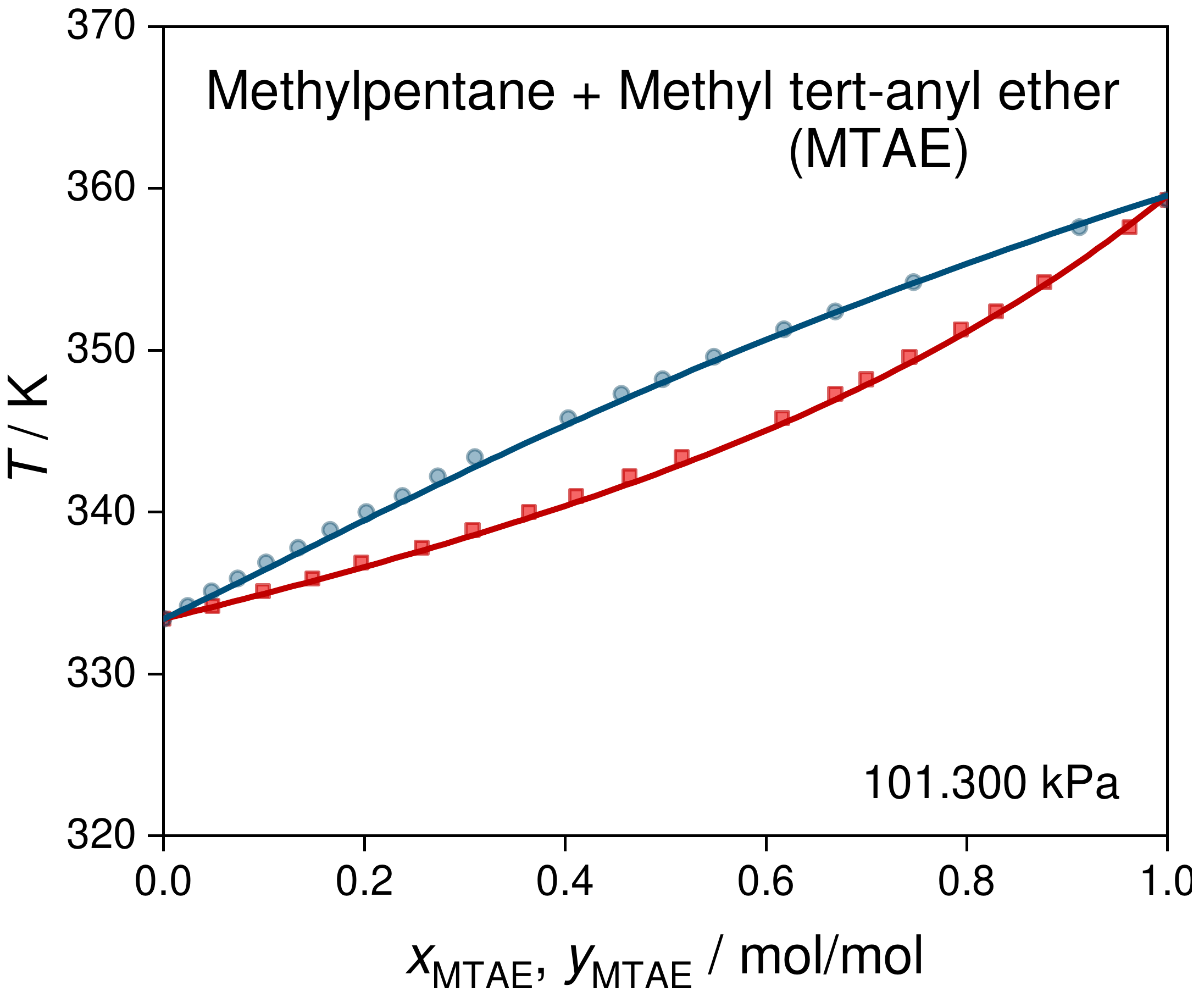}
    \medskip
    \includegraphics[width=0.35\textwidth]{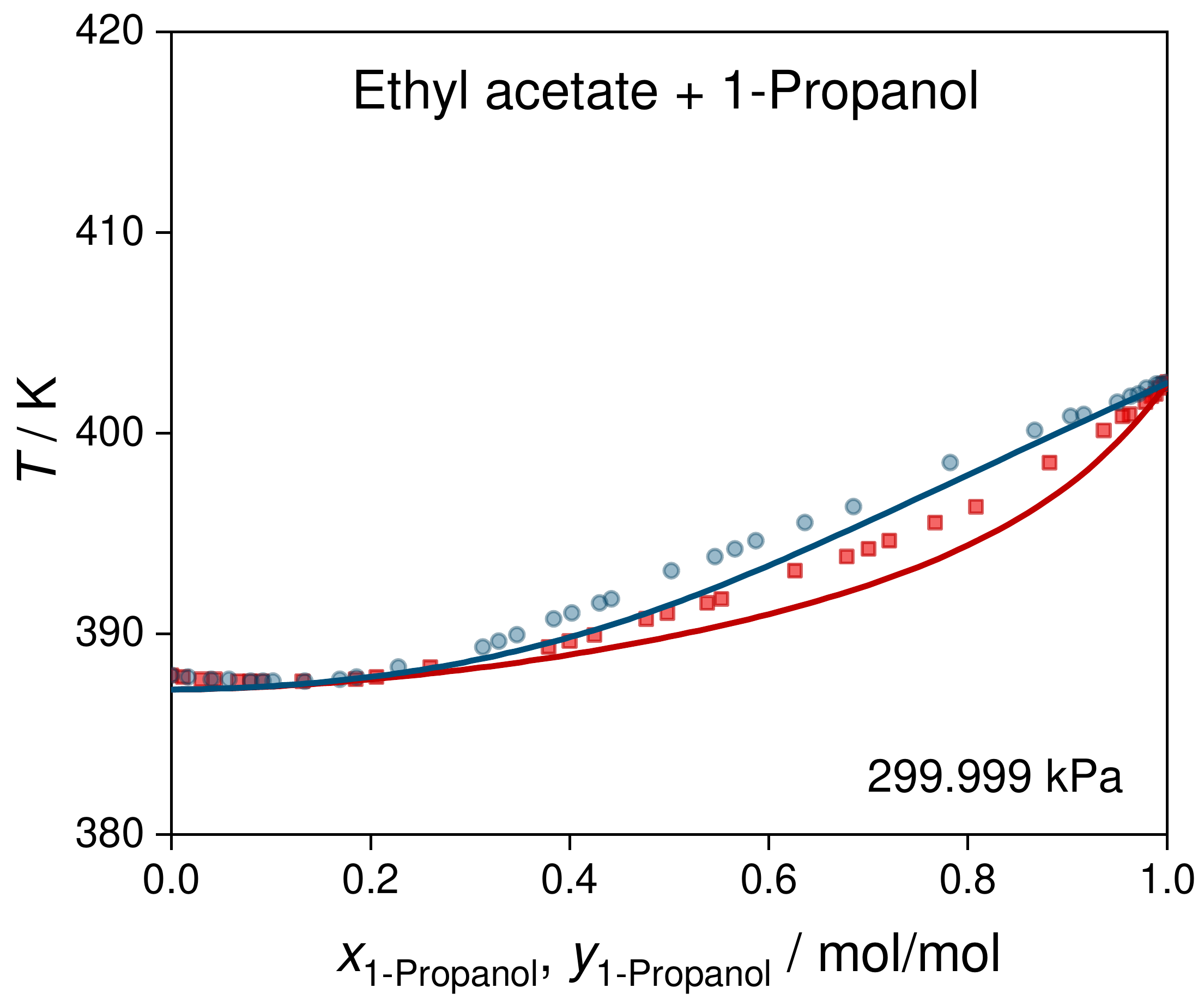}
    \includegraphics[width=0.35\textwidth]{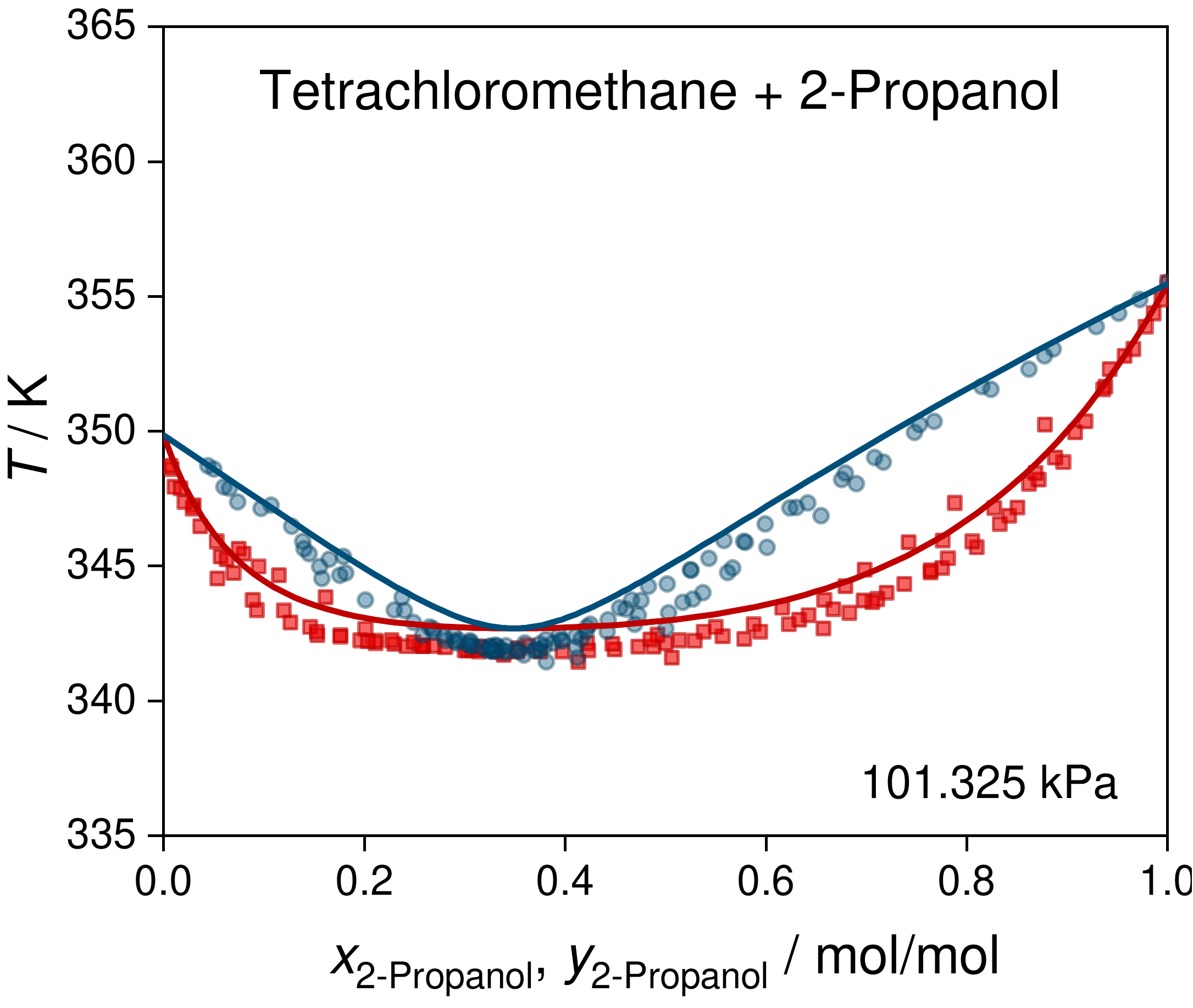}
    \medskip
    \includegraphics[width=0.35\textwidth]{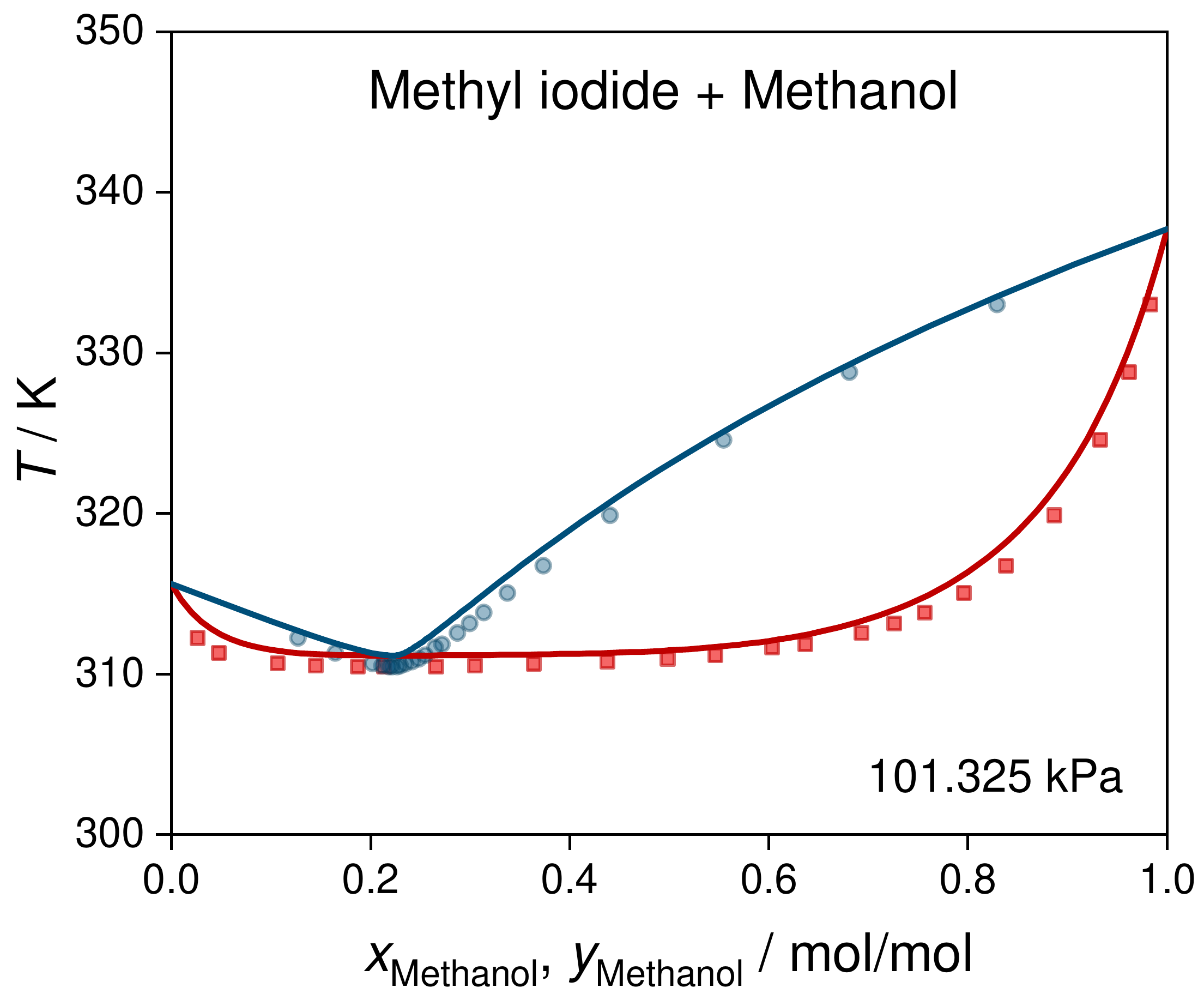}
    \includegraphics[width=0.35\textwidth]{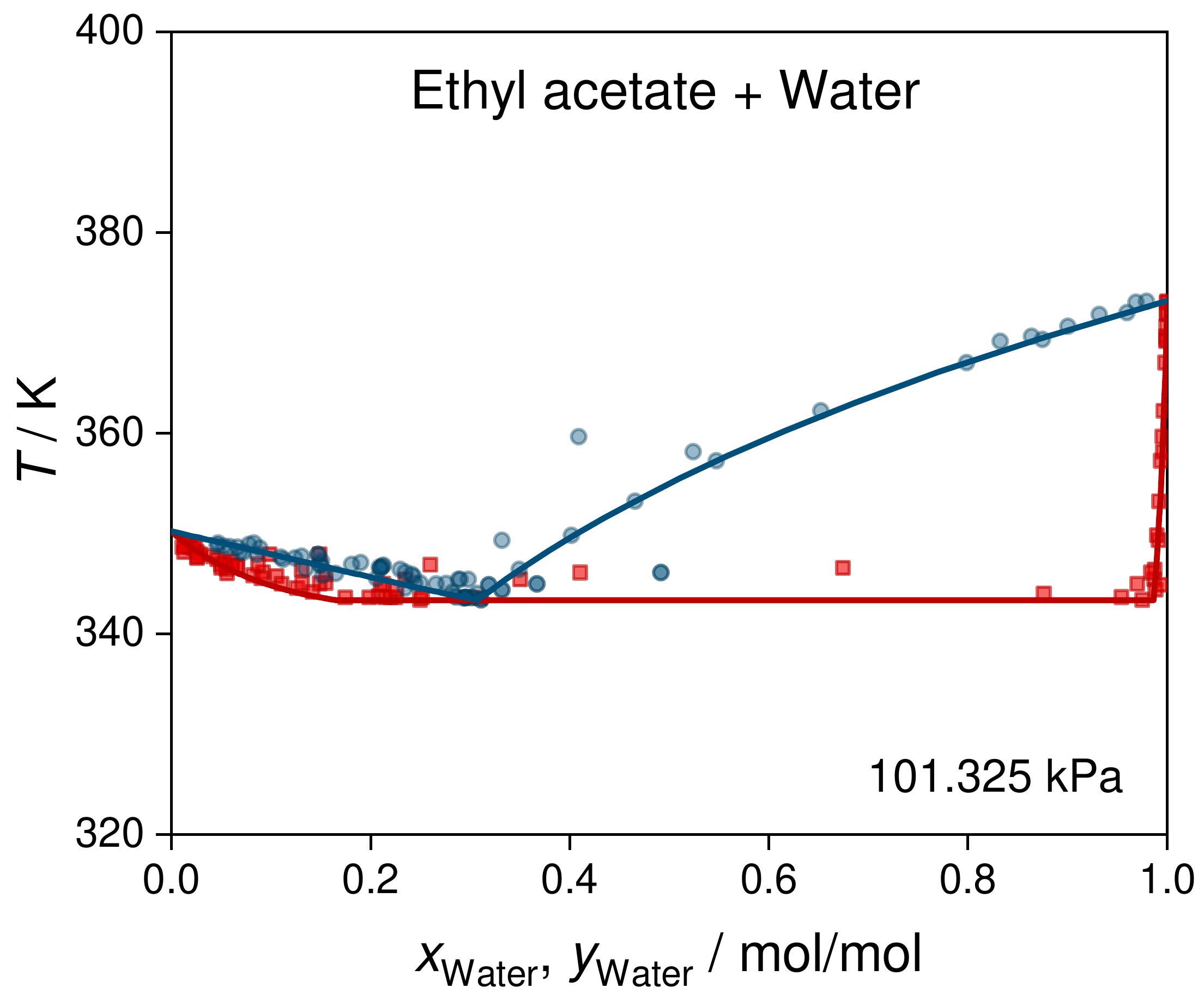}
    \medskip
    \caption{Prediction of isobaric vapor-liquid phase diagrams for binary systems from the test set with MCM-UNIQUAC (lines) and comparison to experimental data from the DDB~\cite{DDB_2020} (symbols). No data on any of the depicted systems were used for training MCM-UNIQUAC or setting the hyperparameters. Blue: dew point curves. Red: bubble point curves.}
    \label{VLE_binary}
\end{figure*}

MCM-UNIQUAC can not only be used for predicting activity coefficients but also for predictions of phase equilibria and many other thermodynamic properties of any mixture consisting of the considered 1,146 components. We demonstrate this in Figure~\ref{VLE_binary}, which shows the results of the prediction of isobaric vapor-liquid equilibrium (VLE) phase diagrams for eight binary systems based on the extended version of Raoult's law, cf. Eq. S.14 in the Supporting Information. All eight systems were chosen randomly from the test set, i.e., not a single data point for these systems was used for training MCM-UNIQUAC or setting its hyperparameters. However, the selection was carried carried out in such a way as to cover a wide range of different phase behaviors, ranging from high-boiling azeotrope (top left) to heteroazeotrope (bottom right). For details on the selection of the systems and the prediction of the VLE phase diagrams with MCM-UNIQUAC, we refer to the Supporting Information.
For all eight binary systems, excellent agreement between the predicted phase diagram and the experimental data is found, both qualitatively and quantitatively. Similarly, liquid-liquid and solid-liquid phase diagrams and other thermodynamic properties like excess enthalpies can be predicted with MCM-UNIQUAC.

\begin{figure}[h]
    \centering
    \begin{subfigure}[c]{\textwidth}
		\centering
    \includegraphics[width=0.5\textwidth]{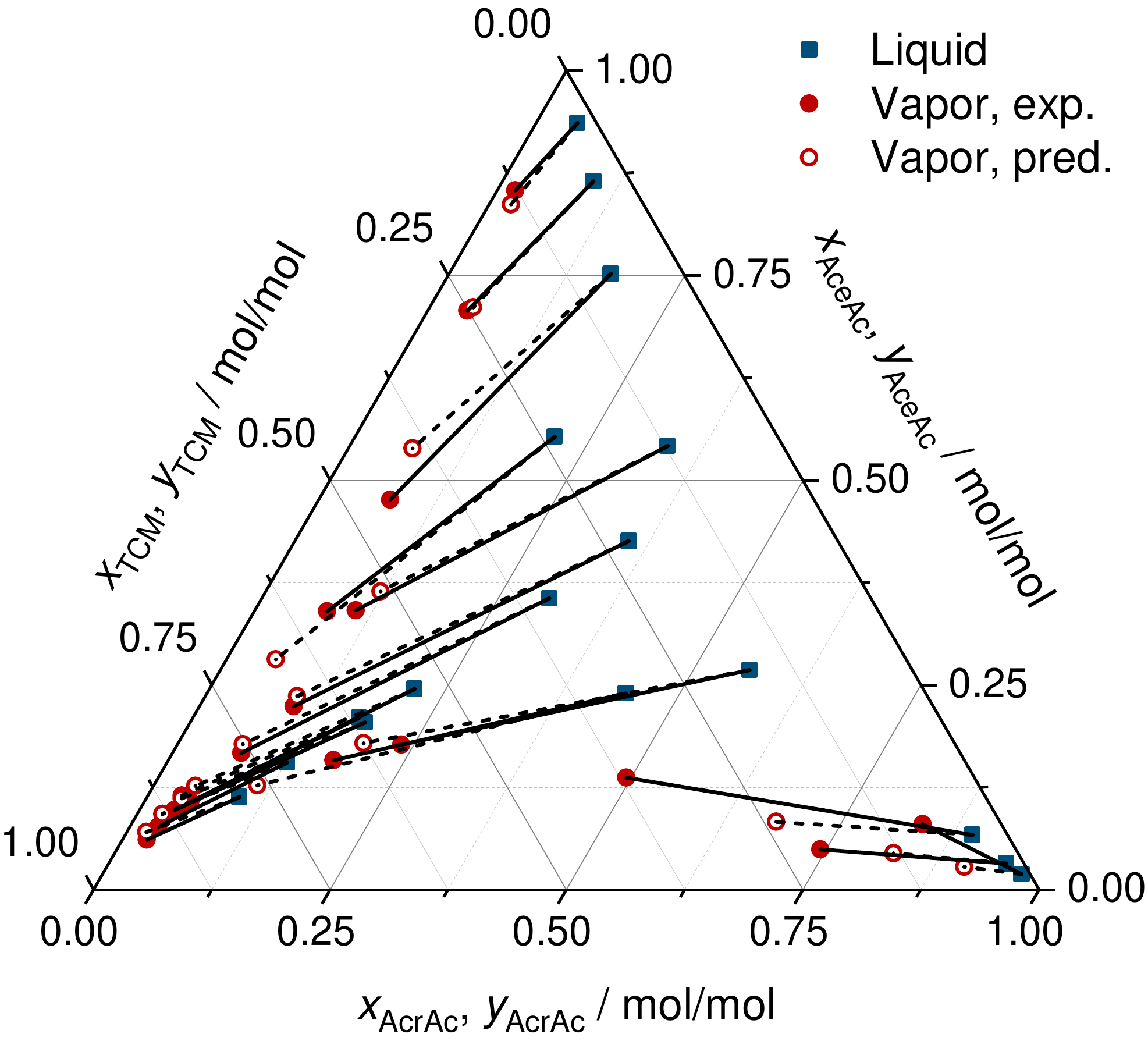}
    \end{subfigure}
    \par\bigskip
    \begin{subfigure}[c]{\textwidth}
		\centering
    \includegraphics[width=0.5\textwidth]{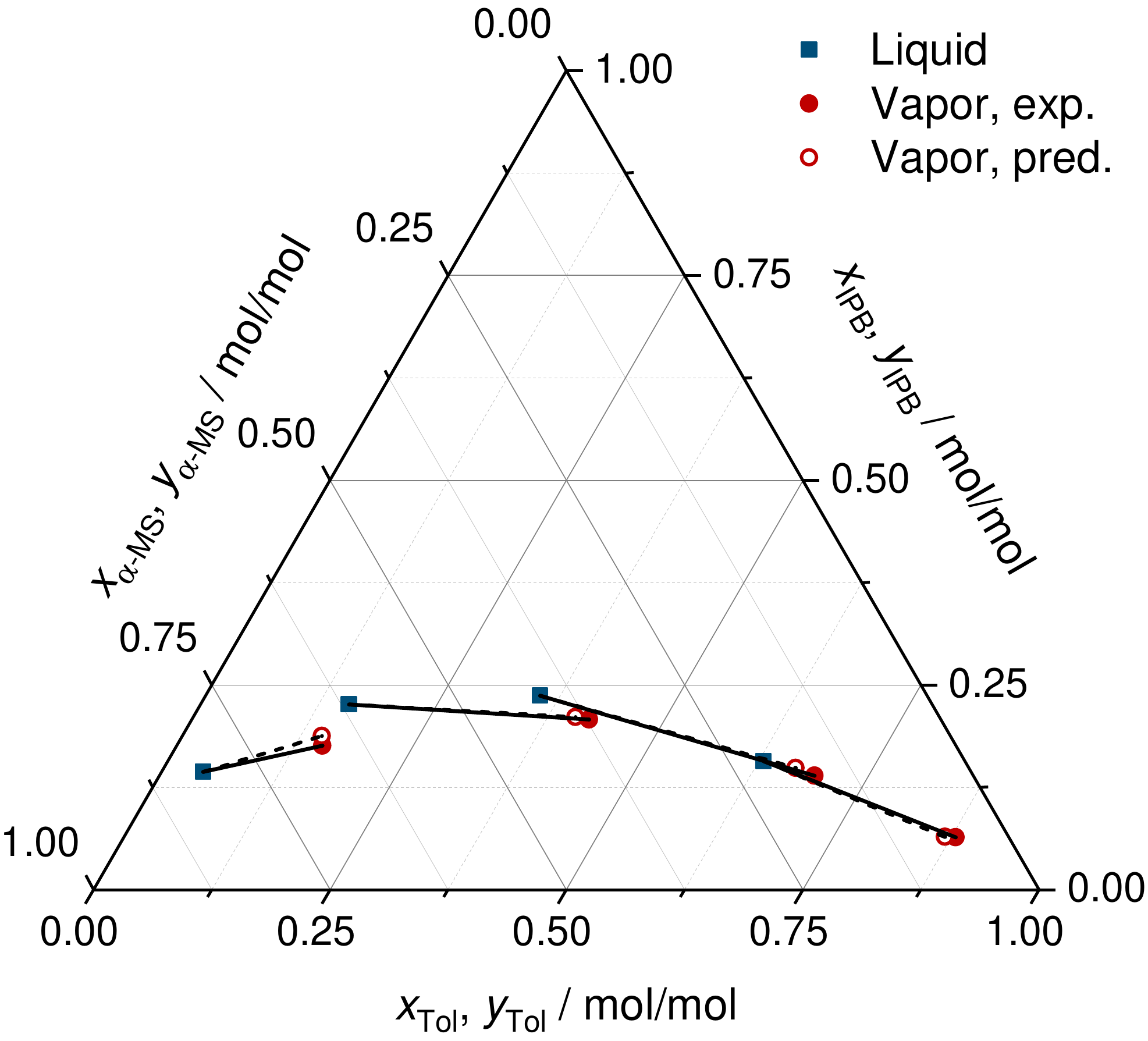}
    \end{subfigure}
    \caption{Prediction of the VLE in ternary systems at constant pressure with MCM-UNIQUAC and comparison to experimental data (exp.) from the DDB~\cite{DDB_2020}. The pressure and the composition of the liquid phase was specified, the composition of the corresponding vapor phase was predicted (pred.). Top: acrylic acid (AcrAc) + acetic acid (AceAc) + tetrachloromethane (TCM) at 100 kPa. Bottom: toluene (Tol) + Isopropylbenzene (IPB) + $\alpha$-methyl styrene ($\alpha$-MS) at 101 kPa. No data on any of the systems and any of the binary subsystems were part of the training set or validation set of MCM-UNIQUAC.}
    \label{VLE_ternary}
\end{figure}

Moreover, the physical foundation on which the hybrid approach MCM-UNIQUAC builds, namely learning and predicting pair-interaction energies between components on the hypothetical lattice from data for binary mixtures, even allows extrapolations from binary to multicomponent mixtures; MCM-UNIQUAC thereby does not require any data of multicomponent systems for training. As an example, isobaric VLE phase diagrams for two ternary systems are shown in Figure~\ref{VLE_ternary}. These systems were selected such that all constituent binary subsystems were neither part of the training set nor of the validation set.

For each data point, the pressure and the composition of the liquid phase (blue symbols in Figure~\ref{VLE_ternary}) were specified. The associated composition of the vapor phase in equilibrium was predicted with MCM-UNIQUAC (open red symbols) and compared to the experimental composition of the vapor phase as reported in the DDB~\cite{DDB_2020} (filled red symbols). We observe an excellent agreement.

In the Supporting Information, we additionally provide a complete set of the pair-interaction energies $U_{ij}$, including the respective model uncertainties in the form of standard deviations, for all 656,085 binary systems of the components considered here (the pair-interaction parameters $\Delta U_{ij}$ of the commonly used version of UNIQUAC can easily be calculated from these by applying Eq.~\ref{eq:delta-u-u}). This set of $U_{ij}$ was obtained from an end-to-end training of MCM-UNIQUAC on the \textit{entire} available data set of all 12,199 systems. If predictions for any system of the components considered in this work are required, this parameter set should be used.

\section{Conclusions}
In the present work, we describe a novel hybrid approach for predicting thermodynamic properties of mixtures, which combines methods from machine learning (ML) with physical modeling. The basic idea is to predict the pair interactions between components in mixtures using matrix completion methods (MCMs). The approach is generic, it can be applied to any mixture property, and any physical model based on pair interactions can be used. As an example, we combined an MCM with UNIQUAC, a well-known lattice model of the Gibbs excess energy $G^\text{E}$. 

We trained our hybrid approach, MCM-UNIQUAC, on experimental data on activity coefficients in binary mixtures of 1,146 components from the Dortmund Data Bank (DDB). Out of the possible 656,085 binary systems that can be formed from these components, suitable experimental data were only available for 12,119 systems, corresponding to only 2\%. In its basic form, UNIQUAC can only be applied to this subset, as pair-interaction parameters need to be fitted to data of the respective systems. In contrast, MCM-UNIQUAC yields, in principle, predictions for all 656,085 binary systems; we demonstrate the predictive capacity of the model based on the systems for which experimental data are available by withholding the test data points during the training of the model. Moreover, by virtue of being a model based on pair interactions, MCM-UNIQUAC can be applied not only to binary systems but also to any multicomponent system that can be formed from the considered 1,146 components; and it also yields predictions for any composition and temperature. 

We compared the quality of these predictions to those from the best available physical model for this purpose, the group-contribution model modified UNIFAC (Dortmund). However, due to missing group-interaction parameters, the public version of UNIFAC can at present only be applied to 9,502 of the 12,119 binary mixtures for which data were available. We show that MCM-UNIQUAC outperforms UNIFAC, even on a test set whose data were not used for training MCM-UNIQUAC, whereas most of these data were probably used for training UNIFAC. 

In its commonly used version, UNIQUAC has two binary parameters $\Delta U_{ij} \neq \Delta U_{ji}$ (with $i \neq j$). It is known that they are highly correlated and hard to interpret physically. In this work, we have discovered that going back to the basic idea of the lattice model and working directly with symmetric pair interactions $U_{ij} = U_{ij}$, i.e., with only a single binary parameter for each binary system, gives almost the same quality of the description of the phase equilibria in the studied binary systems as working with the common version of UNIQUAC with two binary parameters, which is highly remarkable. It can be assumed that the pair-interaction energies $U_{ij}$, which are now available for 656,085 binary systems and 1,146 pure components, can be interpreted physically. As they are basically mean-field energies, this is probably only a first step, and deeper insights may be expected from applying the MCM in connection with more sophisticated mixture models that account for different types of interactions. 

MCM-UNIQUAC combines the strengths of ML and physical modeling - the MCM enables generalization over (discrete) components, allowing predictions for unstudied binary systems, whereas UNIQUAC generalizes over (continuous) conditions and enables the extrapolation to multicomponent mixtures. As demonstrated in examples, MCM-UNIQUAC allows the direct prediction of phase diagrams as exemplified for binary and ternary vapor-liquid phase diagrams. Furthermore, our approach can be retrained easily whenever additional data points become available; this is in contrast to established group-contribution methods, which require extensive parameter tuning. Also, MCM-UNIQUAC does not rely on expensive calculations as required by established quantum-chemical prediction methods. Finally, by exploiting the fact that basically all established thermodynamic models for mixture properties rely on the description of pair interactions, our approach can be further developed in many directions, e.g., by combining MCMs with equations of state and even group-contribution methods.

\section*{Acknowledgments}
The authors gratefully acknowledge financial support by Carl Zeiss Foundation in the frame of the project 'Process Engineering 4.0' and by Bundesministerium f{\"u}r Wirtschaft und Energie (BMWi) in the frame of the project 'KEEN'. This material is in part based upon work supported by the Defense Advanced Research Projects Agency (DARPA) under Contract No. HR001120C0021. Any opinions, findings and conclusions or recommendations expressed in this material are those of the author(s) and do not necessarily reflect the views of DARPA. Stephan Mandt furthermore acknowledges support by the National Science Foundation under the NSF CAREER award 2047418 and Grants 1928718, 2003237, and 2007719, the Department of Energy under grant DE-SC0022331, as well as Intel and Qualcomm.

\bibliographystyle{unsrt}  
\bibliography{MCM2}

\end{document}